# Reputation is required for cooperation to emerge in dynamic networks

Jose A. Cuesta[1,2,3,4], Carlos Gracia-Lázaro[2,4], Yamir Moreno[2,4,5], and Angel Sánchez[1,2,3,4]

[1]Grupo Interdisciplinar de Sistemas Complejos (GISC), Departamento de Matemáticas, Universidad Carlos III de Madrid, Spain. [2]Instituto de Biocomputación y Física de Sistemas Complejos (BIFI), Universidad de Zaragoza, Spain. [3]Institute UC3M-Santander of Financial Big Data (IFiBiD) , Universidad Carlos III de Madrid, Spain. [4]Unidad Mixta Interdisciplinar de Complejidad y Comportamiento Social (UMICCS), UC3M-UV-UZ. [5]ISI Foundation, Turin, Italy.

Melamed, Harrell, and Simpson have recently reported on an experiment which appears to show that cooperation can arise in a dynamic network without reputational knowledge, i.e., purely via dynamics [1]. We believe that their experimental design is actually not testing this, in so far as players do know the last action of their current partners before making a choice on their own next action and subsequently deciding which link to cut. Had the authors given no information at all, the result would be a decline in cooperation as shown in [2].

In [1], the authors carried out a large-scale experiment to study the effect of reputation and clustering on cooperative behavior. Subjects from Amazon's Mechanical Turk were placed on different networks and played an iterated prisoner's dilemma (PD) game [3] with those to whom they were connected. Each participant began with an endowment of 1000 monetary units (MUs) and played 16 rounds. Cooperation consisted in paying 50 MUs, which involved a payoff 100 MUs to its partner, while defection consisted in paying nothing and did not generate any benefit. Accordingly, the possible payoffs were $T=100$, $R=50$, $P=0$, $S=-50$.

The authors considered two kinds of networks, random (Erdös–Rényi) and clustered graphs, with a initial mean connectivity $<k>=4$ in both cases. The clustering coefficients were about 0.167 for the random graphs, and 0.42 for the clustered graphs. In addition, they considered both static and dynamic networks, and three different reputation conditions (no reputation, global reputation and local reputation).

In all dynamic networks treatments, at each round, after playing the PD (1st phase) players were allowed to cut an existing link (2nd phase), to propose a new link (3rd phase) and to accept or not their new link proposals (4th phase). In the 1st and 2nd phases, the information available to players did not depend on the reputation treatment, while in the 3rd and 4th phases did. In the *no-reputation condition*, the authors write that "participants were given no information about potential alters' reputations when adding new ties, which is akin to replacement at random." Nevertheless, as stated in their paper, this information was available in the 1st (PD) and 2nd (link cut) phases: **in the treatments without reputation, subjects knew the last action of each of their partners both when choosing to cooperate or to defect and when cutting a link.** The experiments showed that when people were able to cut and propose new links in all rounds, after 2-3 rounds the level of cooperation was maximum (c~1) regardless of the reputation treatment.

On the basis of their experimental results, the authors conclude that "while reputations are important for partner choice, cooperation levels are driven purely by dynamics". In our study [2], whose setup is closely followed here except that [2] was performed in a controlled (lab) environment, we found that players decided on cutting, proposing or accepting links based on reputation, and that they estimated reputation by combining the last action of the partner and an average of all the available history. When this information was restricted to the last action, they managed links using as reputation the information about the last action of others (or more, if previous round decisions are provided), i.e., it was much more likely to cut a link with a defector than with a cooperator. As a consequence, the fact that actions are known at the time of the link-cutting decision is a strong incentive to cooperate, as otherwise the focal player has every incentive to replace a defector with a random player about whom she has no information. Note that the observation in [2] that a vast majority of subjects cut links with defectors makes very likely that players have this possibility in their minds when choosing their actions. The availability of information on the last action of players at the time on choosing actions and managing links must therefore be interpreted as reputation, and as a consequence the authors are not actually working with a no-reputation treatment.

We thus believe that the experimental results in [1], while being of great interest for instance when comparing choosing the same action for all neighbors with choosing independently for each one, something that has never been compared in the same setup before, do not support the claim that cooperation may emerge without reputation. In this respect, only the setup in [2], with a treatment with no information whatsoever, is a truly non-reputation condition, and in that case the result is a rapid decline in cooperation. In fact, the conclusion in [2] is that reputation is required for cooperation to be stable, but it does not especially promote it when comparing different reputation conditions, a conclusion that is in line with other results on the PD [4], with experiments on the Raub Wessie model [5] and even with the results reported in [1].